# Is Hardy's paradox a paradox?

Y.Ben-Aryeh

Physics Department, Technnion-Israel Institute of Technology, Haifa, 32000, Israel

e-mail: phr65yb@ph.technion.ac.il

**Abstract**

In the present Note it is shown that 'Hardy's thought experiment' does not lead to any paradox and its explanation can be made by using quantum mechanical methods, without the need of 'weak measurements theories'. The confusion arising about this 'thought experiment' follows from ignoring the use of 'a projective measurement of the kind of knowledge'. A certain comment about the realization of 'Hardy's thought experiment' with photons is made.

1. Introduction

A certain 'thought experiment' has been suggested by L.Hardy [1], which can refute local-realism, without the use of Bell inequalities. In various studies which tried to implement and explain experiments of this kind, the experiments have been related to 'weak measurements theories' (see e.g. [2-6]). I would like to show in the present Note that if one uses 'orthodox quantum mechanics' related to 'projection operator of the kind of knowledge' there is not any paradox in this 'thought experiment', and there is no need for using the special theories of 'weak measurements' for explaining *these experiments*. I present a certain comment about the realization of 'Hardy's thought experiment' with photons.

2. A quantum mechanical analysis of 'Hardy's thought experiment'

The Hardy's thought experiment is described in [1] and especially in its Figure 1 as: "two Mach-Zehnder-type interferometers, one for positrons and one for electrons, arranged such that if a positron takes path $u^+$ and electron takes path $u^-$ then they will meet at a point P and annihilate one another".

The input state in this experiment can be described as

$$a_{S_+}^\dagger |0\rangle_{e+} a_{S_-}^\dagger |0\rangle_{e-} = |S_+\rangle_{e+} |S_-\rangle_{e-} \qquad , \qquad (1)$$

where $|S_+\rangle_{e+}$ and $|S_-\rangle_{e-}$ are the input states for positron and electron, respectively. By using the first beam-splitters ($BS1^\pm$) we get the transformations

$$a_{S_+}^\dagger \to \frac{1}{\sqrt{2}}\left(a_{v_+}^\dagger + ia_{u_+}^\dagger\right) \ ; \ a_{S_-}^\dagger \to \frac{1}{\sqrt{2}}\left(a_{v_-}^\dagger + ia_{u_-}^\dagger\right) \quad , \tag{2}$$

where $|v_+\rangle, |u_+\rangle$ are the output states of $BS1^+$ and $|v_-\rangle, |u_-\rangle$ are the output states of $BS1^-$, respectively. In Eqs. (2) we have used a part of the BS's unitary transformations which are relevant to our analysis. One should take into account that in a BS transformation one should take into account two input states and correspondingly two output states (see e.g.[7]) but since in one input port of $BS1^+$ or $BS1^-$ the vacuum is entering and since the vacuum states do not affect our analysis they are omitted for the sake of simplicity.

Using Eqs. (1-2) we find that the input state before the interaction at point P is given by

$$|\psi\rangle = \frac{1}{2}\left\{|v_+\rangle_{e+}|v_-\rangle_{e-} + i|v_+\rangle_{e+}|u_-\rangle_{e-} + i|u_+\rangle_{e+}|v_-\rangle_{e-} - |u_+\rangle_{e+}|u_-\rangle_{e-}\right\} \tag{3}$$

In Hardy's paper [1] it is assumed that if the electron and the positron are in the state $-|u_+\rangle_{e+}|u_-\rangle_{e-}$ then they will interact with absolute probability 1 to produce the state $-|\gamma\rangle$. Then he assumes that the quantum state after the interaction at point P will be given by a certain superposition of quantum states where one of these states will be $|\gamma\rangle$ (See Eq. (9) in [1]). According to my understanding, such assumption is not correct, as a superposition of electron, positron and $\gamma$ states cannot be treated by the effects of the second BS's, i.e., $BS2^\pm$. According to my opinion, we should consider the interaction as 'a measuring projective process of the kind of knowledge'. For a 'measuring process' it is enough if *we know for sure* what will happen at point P, i.e., if we know that at the point P the interaction will produce the photon $\gamma$. The projective measurement in the 'Hardy's thought experiment' produced at point P is then given by

$$\hat{P}_{proj} = \left\{|v_+\rangle_{e+}|v_-\rangle_{e-} + |v_+\rangle_{e+}|u_-\rangle_{e-} + |u_+\rangle_{e+}|v_-\rangle_{e-}\right\}\left\{\langle v_+|_{e+}\langle v_-|_{e-} + \langle v_+|_{e+}\langle u_-|_{e-} + \langle u_+|_{e+}\langle v_-|_{e-}\right\} \quad , \tag{4}$$

and the quantum state after the interaction at point P is given by

$$|\psi'\rangle = C\hat{P}_{proj}|\psi\rangle \quad , \tag{5}$$

where $|\psi\rangle$ is given by Eq. (3) and $C$ is a normalization constant. By straightforward calculations we get :

$$|\psi'\rangle = \frac{1}{\sqrt{3}}\{|v_+\rangle_{e+}|v_-\rangle_{e-} + i|v_+\rangle_{e+}|u_-\rangle_{e-} + i|u_+\rangle_{e+}|v_-\rangle_{e-}\} \quad . \tag{6}$$

By comparing Eq. (6) with Hardy's Eq. (9) of [2], we find two differences: a) In the present analysis the state $|\gamma\rangle$ does not appear in Eq. (6) due to our assumption of 'a projective measurement of the kind of knowledge' which eliminates the amplitude $|u_+\rangle_{e+}|u_-\rangle_{e-}$. b) The state given by (6) has been renormalized. 'Projective measurements of the kind of knowledge' are known from the phenomena of 'which way' (See e.g. [8-9] and references included). 'projective measurement' is not a weak measurement as 'projective measurements of the kind of knowledge' can lead to 'strong' changes in the entangled states [8,9]. Also the simple normalization used in (6) is different from that used in 'weak measurements'.

One should notice that the above projective measurement has led to the state $|\psi'\rangle$ of Eq. (6) which is a pure state due to the assumption that the interaction at point P converted the state $|u_+\rangle_{e+}|u_-\rangle_{e-}$ to a photon $\gamma$ with a probability 1. If we will assume that this interaction occurs with a certain probability less than 1, then a more complicated projective measuring operator should be used which will convert the pure state into a mixed-state, which can be analyzed further only by a density operator formalism. For a comparison between a projective measurement which converts a pure state into a pure state and that of converting a pure state into mixed state see e.g. the discussions in [10]. We will continue the present analysis by using Eq. (6), as the discrepancy between the present approach and that of [1] is manifested already in this simple case:

a) If both $BS2^+$ and $BS2^-$ are removed then

$$|u_+\rangle \to |c_+\rangle \quad ; \quad |v_+\rangle \to |d_+\rangle \quad ; \quad |u_-\rangle \to |c_-\rangle \quad ; \quad |v_-\rangle \to |d_-\rangle \tag{7}$$

and the final state is given by

$$|\psi'\rangle = \frac{1}{\sqrt{3}}\{|d_+\rangle_{e+}|d_-\rangle_{e-} + i|c_+\rangle_{e+}|d_-\rangle_{e-} + i|d_+\rangle_{e+}|c_-\rangle_{e-}\} \quad . \tag{8}$$

b) If $BS2^+$ is in place and $BS2^-$ is removed then

$$|u_-\rangle \to |c_-\rangle \;;\; |v_-\rangle \to |d_-\rangle \;;\; |u_+\rangle \to \frac{1}{\sqrt{2}}\{|c_+\rangle + i|d_+\rangle\} \;;\; |v_+\rangle \to \frac{1}{\sqrt{2}}\{i|c_+\rangle + |d_+\rangle\}, \tag{9}$$

where we have used the BS unitary transformation of $BS2^+$, and the final state is given by

$$|\psi'\rangle = \frac{1}{2}\{|id_+\rangle_{e+}|c_-\rangle_{e-} - |c_+\rangle_{e+}|c_-\rangle_{e-} + 2i|c_+\rangle_{e+}|d_-\rangle_{e-}\} \tag{10}$$

c) If $BS2^+$ is removed and $BS2^-$ is in place then

$$|u_+\rangle \to |c_+\rangle \;;\; |v_+\rangle \to |d_+\rangle \;;\; |u_-\rangle \to \frac{1}{\sqrt{2}}\{|c_-\rangle + i|d_-\rangle\} \;;\; |v_-\rangle \to \frac{1}{\sqrt{2}}\{i|c_-\rangle + |d_-\rangle\} \tag{11},$$

where we have used the BS unitary transformation of $BS2^-$, and the final state is given by

$$|\psi'\rangle = \frac{1}{2}\{|ic_+\rangle_{e+}|d_-\rangle_{e-} - |c_+\rangle_{e+}|c_-\rangle_{e-} + 2i|d_+\rangle_{e+}|c_-\rangle_{e-}\} \tag{12}$$

d) If both $BS2^+$ and $BS2^-$ are in place then

$$|u_\pm\rangle \to \frac{1}{\sqrt{2}}\{|c_\pm\rangle + i|d_\pm\rangle\} \;;\; |v_\pm\rangle \to \frac{1}{\sqrt{2}}\{i|c_\pm\rangle + |d_\pm\rangle\}, \tag{13}$$

where we have used the BS's unitary transformations of $BS2^+$ and $BS2^-$, and the final state is given by

$$|\psi'\rangle = \frac{1}{\sqrt{6}}\{-|d_+\rangle_{e+}|d_-\rangle_{e-} - 3|c_+\rangle_{e+}|c_-\rangle_{e-} + i|d_+\rangle_{e+}|c_-\rangle_{e-} + i|c_+\rangle_{e+}|d_-\rangle_{e-}\} \quad . \tag{14}$$

We notice that a state referred in [2] as a state $|\gamma\rangle$ does not appear in the above $|\psi'\rangle$ final states. The role of the $|\gamma\rangle$ state has not been clarified in [1], while in the present analysis it is eliminated by the use of 'projective measurement'. Since the state $|u_+\rangle_{e+}|u_-\rangle_{e-}$ which appeared in Eq. (3) is eliminated in the reaction process it has been considered as a paradox as the analogous state $|d_+\rangle_{e+}|d_-\rangle_{e-}$ reappear in the final state (14) (As described, for the case where both BS's are in place, by Figure 1 in [1]). However, this result follows from a quantum mechanical analysis and there is not any paradox here. Hardy's thought experiment can, however, be used for refuting local realism, without the use of Bell inequalities [11]. The above analysis has been restricted to the original thought experiment in which the reaction has been assumed to occur with probability 1. It is straightforward to generalize the present approach for cases in which the reaction is known to happen with probability less than 1, by using density operator formalism, but such analysis will be more complicated .

### 3. A comment on realization of 'Hardy's thought experiment' with photons

"Experimental realization of 'Hardy's thought experiment'" has been described in [12]. The experiment consists of a pair of Mach-Zehnder interferometers that "interact" through "photon bunching" in a beam splitter. This bunching effect is, however, different from that of Hardy [1] as the two photons are not annihilated and the bunching effect is related to photon statistics. Therefore, the analog for the projective measurement described in that paper [12] is a post selection measurement with coincidence counters, which follows from conventional quantum optics methods, and there is not any paradox in this analysis, as it has been used only for refuting local-realism.

### 4. Conclusion

Hardy's paradox is not a real paradox as quantum mechanical methods based on 'projective measurement of the kind of knowledge', can be used for analyzing experiments of this type, without the need of using 'weak measurement' theories .

**References**

1. L.Hardy, "Quantum mechanics, local realistic theories, and Lorentz-invariant realistic theories", Phys.Rev.Lett. **68**, 2981-2984 (1992).
2. S.E. Ahnert and M.C. Payne, "Linear optics implementation of weak values in Hardy's paradox", Physical Review A **70**, 042102 (2004).
3. J.S.Lundeen, K.J.Resch, and A.M.Steinberg , "Comment on "Linear optics implementation of weak values in Hardy's paradox",", Physical Review A **72**, 016101 (2006).